\documentclass[12pt,preprint]{aastex}
\lefthead{Reale et al.}
\righthead{Transient loop observed by TRACE I}
\begin{document}
\newcommand{\lpa}{\mbox{L$_1$ }}
\newcommand{\lpb}{\mbox{L$_2$ }}
\newcommand{\lpc}{\mbox{L$_3$ }}
\newcommand{\lpd}{\mbox{L$_4$ }}

\title{The Sun as an X-ray Star: III. Flares}

\author{F. Reale, G. Peres}

\affil{Dip. di Scienze Fisiche \& Astronomiche -- Sez. di
Astronomia -- \\ Univ. di Palermo, Piazza del Parlamento 1, I--90134
Palermo, Italy; reale@astropa.unipa.it, peres@astropa.unipa.it}

\author{S. Orlando}

\affil{Osservatorio Astronomico, Piazza del Parlamento 1, I--90134
Palermo, Italy; orlando@astropa.unipa.it}

\begin{abstract}
In previous works we have developed a method to convert solar X-ray
data, collected with the Yohkoh/SXT, into
templates of stellar coronal observations. Here we apply the method to
several solar flares, for comparison with stellar X-ray flares.  Eight
flares, from weak (GOES class C5.8) to very intense ones (X9) are
selected as representative of the flaring Sun. The emission measure
distribution vs. temperature, EM(T), of the flaring regions is derived
from Yohkoh/SXT observations in the rise, peak and decay of the
flares.  The EM(T) is rather peaked and centered around $T \approx
10^7$ K for most of the time. Typically, it grows during the rise phase
of the flare, and then it decreases and shifts toward lower
temperatures during the decay, more slowly if there is sustained
heating. The most intense flare we studied shows emission measure even
at very high temperature ($T \approx 10^8$ K).  Time-resolved X-ray
spectra both unfiltered and filtered through the instrumental responses
of the non-solar instruments ASCA/SIS and ROSAT/PSPC are then derived.
Synthesized ASCA/SIS and ROSAT/PSPC spectra are generally well fitted
with single thermal components at temperatures close to that of the
EM(T) maximum, albeit two thermal components are needed to fit some
flare decays.  ROSAT/PSPC spectra show that solar flares are in a
two-orders of magnitude flux range ($10^6 - 10^8$ erg cm$^{-2}$
s$^{-1}$) and a narrow PSPC hardness ratio range, however higher than
that of typical non-flaring solar-like stars.

\end{abstract}


\section{Introduction}
\label{sec:intro}

Because of its closeness, the Sun is the best-observed and the only
spatially-resolved X-ray star. It is therefore natural to consider it
as a template and a guide to analyze and interpret what we observe of the
other stars.

In this perspective two previous works (Orlando et al. 2000, Peres et
al.  2000a, hereafter Paper I and Paper II, respectively) illustrate a
method to put solar X-ray data collected by the {\it Soft X-ray
Telescope} on board Yohkoh into the same format and framework as
stellar X-ray data.  The method allows us to simulate accurately the
observation of the Sun at stellar distances with a stellar instrument
and to apply to the relevant data the same analysis as if they were
real stellar data: we can compare homogeneously stellar to solar data
and use the latter as a template for stellar observations.

Paper I focussed on the details of the treatment of Yohkoh data and
Paper II outlined the method in its generality and showed some
representative applications to observations in different phases of the
solar cycle plus one flare case. The present work applies this analysis
to several solar flares in the perspective to interpret stellar flares
and some features of very active stellar coronae.

Coronal flares are transient, X-ray bright and localized events: they
last from few minutes to several hours, they easily overcome the
luminosity of the whole solar corona and they occur in relatively small
regions, often in single coronal magnetic structures (loops).  As such
their occurrence and evolution is mostly independent of the structure
and evolution of the rest of the corona. Since the phenomenology,
duration and intensity can be very different from one flare to another
it is difficult to take a single flare as representative.  In this work
we analyze the emission measure distribution vs. temperature and its
evolution during some selected solar flares, representative of the wide
range of possible events; as we did in Paper II we then use these
EM(T)'s to synthesize relevant stellar-like spectra, which are then
analyzed with standard analysis tools of stellar coronal physics.

Together with the general problem of the structure and heating of the
solar corona, flares represent an unsolved puzzle of Solar Physics.
Their early impulsive phase is so fast that the trigger mechanism
remains elusive, in spite of the high quality of the data collected
with many instruments, and in particular with the Yohkoh/SXT, optimized
to observe flares.  Apart from experimental limitations, intrinsic
physical reasons make the diagnostics of the flare heating very
difficult: the flare starts when a low density plasma is heated at more
than $10^7$ K; due to the low emission measure, initially the plasma is
not easily observable while thermal conduction is so efficient that it
smoothes out in a few seconds any trace of local thermal perturbation.
The rise of the brightness in the soft X-ray band, where flares are
best-observed, does not follow the evolution of the heating (e.g. Reale
\& Peres 1995).

While analysing the rise phase is useful to investigate the
mechanisms which originate the flares, the decay phase has been shown
to be useful in other respects.  It has long been known that the
duration of the decay is linked to the size of the flaring region,
because the thermal conduction cooling time depends on the length of
the flaring loops.  This has direct implications on stellar flares:
from the observed light curve it has been possible to estimate the size
of the unresolved stellar flaring loops. On the other hand it has also
been shown that the estimates based on simple scalings from the
conduction or radiation cooling times only can be largely incorrect:
indeed, significant heating can be present (Jakimiec et al. 1992) and
sustain the solar and stellar flare decays (Reale et al. 1997, Reale \&
Micela 1998, Schmitt and Favata 1999, Favata et al. 2000, Maggio et al.
2000) longer than expected, therefore leading to very coarse loop
lengths overestimates. A useful tool to identify the presence of such a
prolonged heating is the flare density-temperature diagram: the longer
the heating decay (compared to the free cooling loop decay time, Serio
et al. 1991), the smaller is the slope of the decay in this diagram
(Sylwester et al.  1993).  Knowing the decaying light curve and
the path in the density-temperature diagram leads us to obtain reliable
estimates of the flaring loop length and of the heating in the
decay phase (Reale et al. 1997).

The analysis of stellar flares have been often based on loop models
(e.g. Reale et al. 1988, van den Oord et al. 1988, van den Oord \& Mewe
1989, Reale \& Micela 1998). The present study, instead, uses solar
flares really observed and spatially resolved by the Yohkoh/SXT as
templates:  we will apply here the method illustrated in Papers I and
II to obtain the evolving EM(T) and stellar-like flare spectra as
observed by non-solar instruments, in particular the {\it Solid-state
Imaging Spectrometer} (SIS) on board the {\it Advanced Satellite for
Cosmology and Astrophysics} (ASCA) and the {\it Position Sensitive
Proportional Counter} (PSPC) on board the {\it R{\"o}ntgen Satellit} (ROSAT),
which have been among the most used X-ray observatories for the
observation of stellar coronae.

It is interesting to apply the standard stellar spectra analysis to the
resulting flare spectra and see analogies and differences from really
observed stellar flares. The advantages of this approach are that: i)
it is relatively model-independent (only spectral models have to be
included); ii) it analyzes the whole flaring region including
contributions from structures adjacent but outside the dominant (if
any) loop, or, in the case of arcade flares, from many loops.

Spectra of several intense (and therefore yielding good count
statistics) stellar flares collected by ROSAT (e.g.  Preibisch et al.
1993, Ottman et al. 1996) and especially by ASCA (e.g.  Gotthelf et
al.  1994, G\"udel et al. 1999, Osten et al. 2000, Hamaguchi et al.
2000, Tsuboi et al.  2000) have been analyzed in the recent past. In
order to yield a count statistics appropriate for a sound analysis,
each spectrum is typically integrated on time intervals spanning
significant fractions of the flare, therefore averaging on rapidly
evolving plasma and emission conditions. In spite of this, spectrum
models of isothermal plasma (1-T models) are generally able to describe
each spectrum (subtracted by the quiescent "background" spectrum) with
temperatures and emission measures following expected solar-like
trends. The peak temperatures of analyzed stellar flares are quite
higher, and their emission measures orders of magnitude higher ($> 50
MK$ and $> 10^{54}$ cm$^{-3}$, respectively), than typical solar ones.
Deviations from the single temperature description have been also
detected (Ottmann \& Schmitt 1996, Favata et al. 2000). It is
interesting in this context to investigate if there is correspondence
with analogous spectra synthesized from solar flares which may then be
used as templates to interpret stellar flare spectra.

A debated question about stellar flare spectra is the fact that many of
them are better fit by allowing the metal abundance to vary in the
fitting model. The spectra are acceptably fit with metal abundance
increasing in the rising phase and then decreasing gradually in the
decay (Ottman et al. 1996, Favata et al. 2000). The physical meaning of
this result is still far from being settled. Wherever possible, it has
also been shown that the abundance variations seem to change element by
element (Osten et al. 2000, G\"udel et al. 1999).

Although the problem of abundance variations in stellar coronae is
intriguing, far from being settled and addressed by several
investigations with XMM-Newton ({\it X-ray Multi-Mirror Mission}) and
Chandra satellites (e.g. evidence for an inverse First Ionization
Potential effect has been found on the star HR1099, Brinkman et al.
2001), the approach presented here does not allow to investigate it
properly. What we can do is to explore the effect of allowing
abundances to vary in fitting spectra originating from parent
multi-thermal emission measure distributions.

In this study beyond the extrapolation of the solar flares to the
stellar environment as isolated, self-standing and evolving events, we
will also focus on their possible contribution to make up the emission
of very active coronae.  In this respect, evidence has been collected
from multi-line XUV observations that the emission measure of some
active stellar coronae has two peaks, one at a few $10^6$ K and the
other at $\sim 10^7$ K and it has been proposed that the higher
temperature peak is due to a continuous flaring activity (e.g. G\"udel
1997, Drake et al. 2000).  Paper II addressed, although limitedly, the
evidence from extensive ROSAT observations that solar-like stars cover
an extended region in the plane flux/hardness ratio (Schmitt 1997),
while the Sun, as a whole, spans the low HR-low flux part of the same
region. In this context it is worth investigating the region occupied
by flares.

This work is structured as follows: in Section~\ref{sec:data}, we
describe the solar data and our analysis for the derivation of the
flare EM(T)'s and of the stellar-like spectra;
Section~\ref{sec:results} shows the results obtained for our sample of
Yohkoh/SXT flares, the related stellar-like spectra, collected with
ASCA/SIS and ROSAT/PSPC, and their analysis with standard stellar
methods; in Section~\ref{sec:discuss} we discuss our results and draw
our conclusions.

\section{The data analysis}
\label{sec:data}

\subsection{The Yohkoh flare data}

In order to study the EM(T) distribution of solar flares, we have 
selected a sample of flares well-observed for most of their duration by 
Yohkoh/SXT and covering a wide range of flare intensities and physical 
conditions. In particular we have selected eight flares ranging from 
relatively weak (class C5.8) to very intense flares (X9.0). The flares 
and their SXT observations are listed in Table~\ref{tab:fl_list}, along 
with the date and time of the flare beginning as measured with the {\it
Geosynchronous Operational Environmental Satellite} (GOES), the 
GOES class, the duration of the flare as in the GOES log file, and the 
start, maximum emission measure and end times of the SXT observations.  
The observations monitor large fractions of the rise, peak and decay of 
the thermal phase of the flares.  

The analyzed data include the SXT data taken in flare mode, and, in
particular, in the two filter passbands specific for flare mode
observations, i.e. Be 119 $\mu$m and Al 11.4 $\mu$m. As discussed in
Section \ref{sec:emt}, additional data taken in two other softer
filters, Al.1 and AlMg, are included to analyze flare 4.  The SXT data
have been processed according to the standard Yohkoh Analysis System.
The datasets consist of sequences of frames $64\times64$ pixels of 2.5
arcsec side, taken alternatively in the two filter bands, with sampling
cadence usually ranging between $\sim 10$ sec (typically
in the rise phase and around the flare peak) and $\sim 60$ sec in
late decay phase of long-lasting flares.

The analysis then includes the derivation of temperature and EM maps,
as in Papers I and II, during the flare evolution. Since the plasma
conditions may change significantly in the time taken to switch between 
the two filters, , the data in the Al 11.4 $\mu$m filter band have been 
interpolated to the exact times of the Be 119 $\mu$m data (as done 
routinely in the standard Yohkoh data analysis), in order to improve the 
accuracy of the temperature and emission measure estimates.  

The distributions of EM vs T (EM(T)) are then derived from the T and EM
maps with the procedure illustrated in Paper I and II. In any of the
two filter bands we have screened out saturated pixels, typically in the 
brightest regions, and pixels which collected photons below a threshold 
of 10 photons.  These pixels would introduce large errors and are 
therefore critical in the analysis of EM vs T for localized events like 
flares. We have carefully selected for our analysis only those frames 
with no (or just very few) saturated pixels.  The temperature bins for 
the EM(T) are those defined in Papers I and II, i.e. 29 bins uniform in 
$\log T$ between log T = 5.5 and log T = 8.  

\subsection{From the EM(T) distribution to the stellar-like spectra}
\label{sec:spec}

From each of the several EM$(T)$ distributions of a flare obtained
during its evolution we synthesize the relevant X-ray spectrum with the
MEKAL spectral code (Kaastra 1992; Mewe et al. 1995), and filter it
through the instrumental response of non-solar telescopes of interest. The
process to generate the stellar-like spectra from an EM$(T)$
distribution derived from the solar data is described in Paper II.
Metal abundances are assumed as in Anders \& Grevesse (1989).

In this paper, we consider the stellar-like focal plane spectra that 
{\em ASCA}/SIS and {\em ROSAT}/PSPC  would collect, observing the solar 
flares selected here at stellar distances.  It would be very interesting 
to perform finely time-resolved spectroscopy, but in real stellar 
observations this is prevented by the limited counting statistics of the 
source. Analogously to stellar observations, therefore, the flare 
spectra have to be integrated on time intervals of hundreds of seconds 
to increase the overall statistics of the fitting process. This means 
that the flares are binned into few time intervals.  Since flares evolve 
on time scales smaller or of the same order, the data collected in a 
time bin do not represent steady plasma conditions but integrate on 
significant variations of the plasma temperature and density in the same 
dataset.  Therefore when time-resolved flare data are analyzed with 
steady-state models, one should keep anyhow in mind that this is an {\it 
a priori} limited representation and description of the data. In order 
to approach the typical conditions of non-solar flare observations we 
make the following exercise: 

\begin{enumerate}

\item we sample the EM(T)'s of a flare at constant time intervals of 60 sec

\item from each EM(T) we synthesize the corresponding spectrum filtered 
through the spectral response and effective area of two instruments,
namely ASCA/SIS and ROSAT/PSPC

\item we bin the flare into three (or more) long time segments, (at
least) one including the rise phase, one the flare maximum and one the
decay phase. The duration of the time segments varies within each flare
and from flare to flare, depending on the flare duration and on the
data structure (the presence of gaps, for instance), and span from a
minimum of 180 s to a maximum of $\sim 3000$ s.

\item each spectrum is normalized so as to yield a total number of
counts in each time interval typical of good stellar flare observations
(e.g. between 1000 and 10000 counts for ASCA);  to this end we have to
assume the distance at which the solar flare would yield the
appropriate statistics.  We randomize the photon counts according to
Poisson statistics

\item we sum all folded and randomized spectra within each time
bin and therefore obtain a single spectrum for each time bin. 

\item we analyze each spectrum with the standard thermal models used for 
stellar data analysis.  

\end{enumerate}

We apply the standard stellar analysis to the binned spectra and use
the tools commonly used by the stellar community. In particular, we fit
the stellar-like spectra with multiple-isothermal components models. In
the ROSAT/PSPC case, we exclude channels with less than three counts
from the analysis to grant an appropriate evaluation of $\chi^2$, and
the first two channels because they are typically affected by
systematic errors.  In the ASCA/SIS case, the energy channels are
grouped so to have at least 20 photon counts per channel; furthermore
the channels with bad quality or empty are discarded\footnote{The first
18 channels of ASCA/SIS below 0.5 keV are discarded because they are
affected by systematic errors.}.  We use the X-ray spectral fitting
package XSPEC V10.0 and the manipulation task FTOOLS V4.0. 
Most fittings are performed with single temperature components, assuming
negligible column density $N_H = 0$, and standard metal
abundances kept fixed. Some checks have been done by allowing metal
abundance to vary all by the same amount.

\section{Results}
\label{sec:results}

\subsection{Flare properties}

All flares listed in Table~\ref{tab:fl_list} last between 9 and 90 min,
except the huge flare 8, of class X9, whose duration is more than 6 hours
(despite the GOES log reports less than 1 hour).  The SXT observation of 
flare 8 is divided into 3 segments separated by two gaps lasting more 
than one and two hours, respectively.  The first segment begins just 
around the flare maximum, and therefore most of the flare rise phase is 
not covered by SXT. We have selected anyhow this flare because it is a 
rather extreme case of intense flare and the information from the rise 
phase is not crucial for our findings, as described below.  

\placefigure{fig:lcnt}
\placefigure{fig:lcnt2}

In all the other cases the related SXT observation covers quite well
both the rise and the decay phase and the flare maximum is well within
the observation. Only for flare 7 (class X1.5) the decay phase is
monitored for a relatively short time. 

As part of our analysis we investigate the role of the morphology of the 
flaring region in determining the EM(T), and the effect of the heating 
release, and in particular of its intensity, duration and decay time, on 
the evolution of the EM(T). In this perspective we tag each flare with 
the light curve in both filters of SXT flare-mode, its path in the 
density-temperature (hereafter {\it n-T}) diagram (the square root of 
the emission measure is used as proxy for the density), and the main 
morphological features observable in the SXT images.  
Fig.~\ref{fig:lcnt} and Fig.~\ref{fig:lcnt2} show the light curves in
the Al 11.4 $\mu$m and Be 119 $\mu$m filter bands obtained by summing
all the counts in each 64x64 pixels SXT frame, and the corresponding
n-T diagrams, obtained from the ratio of the light curves data points.
Notice that the resulting temperature is a weighted average temperature
of the whole region in each frame.

Flare 2 is particularly well-covered since the total luminosity at the 
end has decayed to the values at the beginning; this corresponds to a 
closed cycle in the n-T diagram. The n-T diagram also shows that this is 
the only flare in which the temperature changes significantly  during 
the rise phase (from $\log T \approx 6.8$ to $\log T \approx 7.1$). This 
may indicate that the heating which triggers the flare is released more 
gradually in the rise phase of this flare than in the others (e.g. 
Sylwester et al. 1993). During the other six flares for which the rise 
phase has been observed, the temperature is in fact much more constant 
and stays above $10^7$ K during the rise phase. As for the density, 
Yohkoh/SXT has detected an increase of more than half a decade of 
EM$^{1/2}$ for five flares (flare 2, 4, 5, 6 and 7); during the decay of 
flare 8 the decrease of EM$^{1/2}$ is particularly significant, i.e.  
almost one order of magnitude.  

Table~\ref{tab:fl_par} shows relevant physical and morphological
characteristics of the selected flares obtained with the flare mode 
filters, i.e. the slope of the flare decay path in the $\log(EM^{1/2})-
\log(T)$ diagram, the decay time of the light curve in the Al 11.4
$\mu$m filter band (the latter two quantities are used to estimate the
flaring loop length according to Reale et al. 1997), conservative loop
half-length ranges obtained from measuring the loop projections on the
images and from applying the method of Reale et al. (1997), the
morphology of the flaring region (Fig.~\ref{fig:frames} shows one
grey-scale image sampled during each flare), the maximum temperature and 
emission measure obtained with the filter ratio method from the ratio of 
the data points of the light curves in the two flare-mode filter bands.  
The slope $\zeta$ is an indicator of sustained heating during the decay, 
whenever significantly smaller than $\sim 1.7$ (Sylwester et al. 1993).  

\placefigure{fig:frames}

From Table~\ref{tab:fl_par} we see that:
\begin{itemize}

\item Heating is negligible during the decay of three flares (1, 3 and
6), significant during the decay of the other five and in particular of
the most intense ones (7 and 8). The slope $\zeta$ of the whole decay
of flare 8 is below the minimum possible value ($\zeta_{min} \approx
0.3$) predicted by single loop hydrodynamic modeling (Reale et al.
1997), which means that this model is not applicable in this case, and
that, therefore, significant magnetic rearrangements and complicated
and continued heating release probably occur.

\item The light curve e-folding decay time is below 30 min for all
flares except flare 4, an arcade flare, and flare 8.

\item The loop half-length is in the range 10 Mm and 100 Mm,
typical of active region loops\footnote{Since flare 4 is an arcade 
flare, it involves bundles of loops and the loop length in
Table~\ref{tab:fl_par} is to be taken as an indicative scale size.}.

\item The morphology of the flaring regions ranges from simple single
loops to multiple loops, an arcade and even more complicated structures.

\item The maximum values of the average temperatures, as measured with
Yohkoh filter-band ratio, are in the range $7.1 \leq \log(T) \leq 7.3$,
weakly increasing with the GOES class.

\item The maximum total emission measure increases with the flare
GOES class and spans two orders of magnitudes from $10^{49}$ to 
$10^{51}$ cm$^{-3}$.  

\end{itemize}

The above considerations make us confident that this sample of 
flares is enough representative of flare conditions typically met on 
the Sun and can be used to derive general properties to be compared to 
those observed in stellar flares.  

\subsection{Flare EM(T)}
\label{sec:emt}

Fig.~\ref{fig:emt} shows the evolution of the EM(T) obtained with the
two hardest SXT filters, during the eight selected flares \footnote{For
the sake of clarity, the flare EM(T)'s are not shown as histograms.}.
For each flare we show the emission measure distributions sampled at a
constant rate of one every 2 min since the beginning of the data
selected as in Section \ref{sec:data}.  The longer the flare, the more
are the EM(T)'s shown: those shown for the short flares 1 and 7 are
much fewer than those shown for flare 8, by far the longest one. For
reference, Fig.~\ref{fig:emt} shows also the EM(T)'s obtained in Paper
II for the Sun near the maximum and the Sun near the minimum of the
cycle.

\placefigure{fig:emt}

Fig.~\ref{fig:emt} shows that at any flare phase the EM(T) is typically 
quite narrow, practically independent of the flaring region morphology, 
and covers a temperature decade around $\sim 10^7$ K. It is much 
narrower than any EM(T) of the whole Sun. Exceptions are the EM(T)'s
obtained for flare 8, the most intense one, which shows significant 
amounts of hotter plasma, at temperatures up to $\sim 10^8$ K. The EM(T) 
reaches for this flare maximum values $> 10^{50}$ cm$^{-3}$. Hot plasma 
at temperatures above 30 MK is present in flares more intense than M1.1 
(flare 3).  

In Fig.~\ref{fig:emt} the EM(T) of all flares clearly follows a common 
evolution path: it starts low but already at a relatively high 
temperature, centered at $\sim 10^7$ K; it grows toward higher EM 
values, maintaining a more or less constant shape and sometimes shifting 
slightly rightwards to higher temperatures (e.g. flares 2, 4 and 6); 
then it decays by gradually cooling (leftwards) and decreasing 
(downwards). We can clearly identify envelopes of the evolving EM(T)'s 
and notice that the slope of the envelopes in the decay phase 
(determined by the relative rate of the emission measure decrease and
the temperature decrease) is linked to the slope in the n-T diagram
(see Fig.~\ref{fig:lcnt} and Fig.~\ref{fig:lcnt2}). Several EM(T)
distributions late during the decay are partially hidden by the higher
preceding ones (e.g.  flares 3, 4 and 5). This clearly shows that the
EM(T) mostly evolves inside a ``common envelope" in the rise and the
decay phase, with important implications on the interpretation of some
stellar coronal EM(T)'s (see Section~\ref{sec:discuss}).

The EM(T)'s shown in Fig.~\ref{fig:emt} are derived from observations in 
the hardest SXT filters bands. Since these filters are most sensitive to 
plasma around and above $10^7$ K, one may wonder whether contributions 
from plasma at lower temperatures, not detected by the two filters, may 
be important to the flare EM(T), or not.  The observation of flare 4 
includes several frames taken in the Al.1 and AlMg filter bands and 
allows us to investigate this item.  Fig.~\ref{fig:emt_4} shows the 
EM(T)'s of flare 4 at the same times as those shown in 
Fig.\ref{fig:emt}, but including contributions to the EM(T)'s derived 
from images in the softer filters bands mentioned above.  Cooler 
contributions to the EM(T)'s are clearly present from comparison with 
Fig.\ref{fig:emt}. These contributions make each EM(T) flatter at 
temperatures below the EM(T) maxima ($\sim 10^7$ K); the slope in this 
rising region approaches the $T^{3/2}$ trend expected from loop 
structures (see Peres et al. 2001).  

Although the presence of such contributions modify the shape of the 
EM(T)'s in the low temperature part, they are anyhow significantly 
smaller than the dominant components around $10^7$ K, and we have 
checked that the analysis of stellar spectra synthesized from EM(T)'s 
with and without such contributions does not change significantly, as 
further discussed in Section~\ref{sec:spectra}.  Since very few flares 
yielded good data in softer filters bands (saturated pixels are much 
more frequent), and contributions from such bands are anyhow affected by 
uncertainties and contaminations from the coexistence of cool and hot 
plasma within the same pixel, we prefer to perform the analysis 
considering expositions only in the two hardest filters and keeping 
anyhow in mind the limitations that this choice implies.  

\placefigure{fig:emt_4}

All EM(T)'s obtained for flares are well separated and distinct from
the EM(T) of the Sun at minimum. They are also mostly ``higher'' than
the EM(T) of the Sun at minimum. The EM(T)'s of the first three flares
are instead all lower than the EM(T) of the Sun at maximum, while the
EM(T)'s of the remaining five flares are comparable or higher. If we
did the exercise, similar to that in Paper II, to build a single EM(T)
of the flaring Sun at maximum by combining that of the Sun at maximum
and any one of the latter flares, we would invariably obtain an EM(T) 
with two distinct peaks, the hotter one being associated with the flare. 
Flare 8 involves emission measures quite higher than the emission 
measure of the full non-flaring Sun at maximum (see Paper II).  

Fig.~\ref{fig:maxemt} shows the maxima of the EM(T)'s of all of the
eight flares.  This figure indicates that an increasing intensity in
the GOES class mostly corresponds to an increase in the EM(T) height
and much less to an increase in temperature and/or in the EM(T) width.
An exception is flare 8 which is significantly hotter and wider.

\placefigure{fig:maxemt}

\subsection{Flare spectra}
\label{sec:spectra}

From each EM(T) obtained during the flares we can synthesize
the corresponding X-ray parent spectra for an optically thin plasma in
thermal equilibrium, as described in Paper II.  Fig.~\ref{fig:spectra}
shows examples of X-ray spectra in the 0.2-20 keV band, synthesized
from EM(T) at the beginning, peak and end of Flare 2 and at the peak
and data end of Flare 8.  The maximum luminosities in the whole band
result to be $\approx 3.5 \times 10^{26}$ erg/s and $\approx 1.3 \times 
10^{28}$ erg/s, respectively. From Fig.~\ref{fig:spectra} we see 
concentrations of emission lines around 1~keV, typical of plasma mostly 
at temperature around 10$^7$ K, and mostly due to the Fe-L complex. The 
spectrum at the peak of flare 8 shows prominent lines of the Fe group at 
energies around 6 keV, which are sensibly reduced at the end of the data 
(lower panel, dashed line). Notice that the continuum of the spectrum at 
the peak of flare 8 is considerably flatter than the others shown and 
indicates that there are contributions of plasma at significantly hotter 
temperatures than in all other synthesized spectra shown.

\placefigure{fig:spectra}

\subsubsection{ASCA/SIS spectra}
\label{sec:asca}

Fig.~\ref{fig:spectra} shows ASCA/SIS spectra of flare 4.  These spectra 
have been analyzed by fitting them with spectral models consisting of 
one isothermal plasma component ({\it 1-T fit}).  
Table~\ref{tab:fl_asca} shows the results of the spectral analysis 
applied to four of the eight selected flares (2, 4, 6, and 8).  For each 
flare, the Table includes the distance at which we put the flaring Sun, 
and, for each time bin, the phase of the flare it covers, the time 
range, the number of counts, the best-fit temperature and emission 
measure, the number of degrees of freedom (the number of energy channels 
yielding a significant number of counts minus the parameters of the 
fitting) and the reduced $\chi^2$ of the fitting.  

\placefigure{fig:spectra}

Table~\ref{tab:fl_asca} tells us, first of all, that the single thermal
component model already provides an acceptable description of the 
various phases of the flares, in agreement to the rather peaked EM(T) 
distributions that we obtain from the two hard SXT filters, even 
averaging over each time bin. In spite of the fast flare evolution, 
temperature and emission measure variations {\it within} each time bin 
do not affect the 1-T fitting in the rise phase, mainly for two reasons:  
i) the emission measure at the end of the bin in the rise phase is much 
higher than, and dominates over, that at the beginning; ii) generally 
the temperature does not vary much during the rise phase. Conditions are 
quite stable during the flare maximum and a single temperature therefore 
describes reasonably well this phase.  

On the contrary, in spite of the relatively slow evolution, during the 
decay plasma temperature may vary significantly from the beginning to 
the end of the same time bin, while the emission measure remains 
relatively constant (or slowly decreasing). The single temperature 
component may therefore fail to describe time bins of the decay phase, 
especially in flares with a steep slope in the n-T diagram.  Indeed in 
bins e) and f) in the decay phase of flare 6, the fitting is not as good 
($\chi^2 > 1.7$) as in the other ones or in the other flare decays: the 
slope of the decay of this flare in the n-T diagram is very high ($\zeta 
> 1.7$), higher than in all the other flares. 1-T fitting of spectra 
taken with ASCA during stellar flare decays, and subtracted of the 
spectrum of the background non-flaring corona, have sometimes failed 
(e.g.  Favata et al. 2000), and our results suggest a possible 
explanation.  

The best-fit temperature values approximately correspond to the maxima 
of the EM(T) distribution averaged in the respective time bin and the 
corresponding emission measure values are approximately proportional to 
the total emission measure in that bin. The maximum emission measure 
values are slightly smaller than the values listed in 
Table~\ref{tab:fl_par}, both because some contributions are excluded by 
the instrument limited spectral band and because in the respective time 
bins of Table~\ref{tab:fl_asca} the emission measure is not constantly 
at maximum. Notice that due to scaling with respect to the maximum, the 
last bin of flare 8 yields only 400 photon counts, since the emission 
measure is two orders of magnitude smaller than at maximum.  

The last section of Table~\ref{tab:fl_asca} reports again results for 
flare 4 but including the components obtained from the softer filters
(see Fig.\ref{fig:emt_4}). Comparing to the results without the softer 
filters, we notice the slightly higher count statistics, the slightly 
lower best-fit temperature, the slightly larger emission measure, the 
generally higher $\chi^2$. The higher $\chi^2$ indicates that the single 
thermal model is not as good as previously, but not to such a point to 
discriminate the presence of the cooler components, also considering the 
higher count statistics.  

Finally, we have checked the effect of fitting the simulated ASCA data
with 1-T models in which the global metal abundance is left free to
vary, from one spectrum to the other. The resulting best-fit abundance
values are generally different from the expected solar value and vary
within a factor two, either in excess or in defect, from the central
unity value. In the course of a flare we obtain either abundances all
smaller than one, or all larger than one, or fluctuating around one.
Although in a few cases the abundance appears to decrease during the
decay, we were not able to identify clear systematic trends of
abundance variations during the flare evolution. The best-fit $\chi^2$
values are improved by 0-10\%, and in no case by more than 20\%, while
the temperature values are practically unchanged (while the emission
measure values vary in inverse proportion to the abundance values).
The detection of such abundance variations are clearly an artifact of
the model fitting process, since the synthesized spectra are
consistently built by assuming fixed solar metal abundances at any step
and can be explained in the following terms:  the EM(T) flare
distributions are neither sharp enough to be described at best by a
simple $\delta$ function (an isothermal model), nor broad enough to
require a multi-thermal (e.g. 2-T) model. Moderate abundance variations
are sufficient to account, and adjust the fit, for the presence of
minor emission measure components around the dominating maximum
component.  This exercise tells us that although the limited broadness
of the flare EM(T) may favor an improvement of spectral fitting by
letting abundance vary, the resulting abundance variations seem to be
random; therefore we cannot exclude that systematic abundance
variations are to be explained by effects other than the one
illustrated here.

\subsubsection{ROSAT/PSPC spectra}

ROSAT/PSPC has a lower spectral resolution than ASCA/SIS, and
is, therefore, less able to discriminate multi-thermal components.
In fact, we find that a single thermal component model describes even
better the PSPC spectra obtained from the flare EM(T)'s averaged over
the same time bins as those used for ASCA/SIS. We have analysed spectra
yielding a maximum between 3000 and 5000 total counts:  all fittings
were acceptable (reduced $\chi^2<1.5$ with less than 30 degrees of
freedom).  The temperature and emission measure values are very similar
(with a mean deviation within 10\% and 15\%, respectively) to those
obtained with ASCA/SIS spectral fitting. On the average, the ROSAT
temperature is slightly systematically lower ($\sim 3$\%) than the
corresponding ASCA one, and the emission measure slightly higher ($\sim
1$\%).

Analogously to Paper II, as an additional piece of analysis of the 
synthesized ROSAT/PSPC data, we compute the surface X-ray flux $F_X$, 
defined as the X-ray luminosity in the ROSAT/PSPC spectral band divided 
by the pixel area over which the Yohkoh/SXT photon count is larger than 
10 cts/s, and the hardness ratio defined as in Schmitt (1997): 

\begin{equation}
HR = \frac{H-S}{H+S}
\label{hr}
\end{equation}

\noindent
where $S$ are the total counts in the (soft) PSPC sub-band 0.13-0.4 keV
and $H$ are the total counts in the (hard) sub-band 0.55-1.95 keV.

Fig.~\ref{fig:rosat_flux} shows $F_X$ vs HR for each time bin in which
a spectrum has been collected, and for all flares. All data points
appear to lie in a relatively thin vertical strip around a hardness
ratio value of 0.3 and with $F_X$ ranging from $10^6$ to $10^9$ erg
cm$^{-2}$ s$^{-1}$.  The smallest flares evolve mostly in the range
$10^7$ to $10^8$ erg cm$^{-2}$ s$^{-1}$, while the most intense ones 
seem to move in a wider flux range.  The flare evolution has no well 
defined trend in this plane, mostly because of the very limited 
variation of the hardness ratio. The almost constant HR is mostly due to 
two factors: i) the flare average temperature, of which the hardness 
ratio is a proxy, does not change much from $T \sim 10^7$ K, during the 
flare evolution; ii) the hardness ratio, as defined above, is weakly 
sensitive to temperature variations (and even multi-valued) around $T 
\sim 10^7$ K. For comparison, Fig.~\ref{fig:rosat_flux} shows the plane 
region occupied by late-type stars observed by ROSAT/PSPC (Schmitt 
1997): solar flares (isolated from the remaining corona), practically in 
any phase of their evolution, are harder than typical stellar coronal 
emission and in a region of high surface flux.  

\section{Discussion}
\label{sec:discuss}

Papers I and II illustrated a method to use the Sun as a template of
X-ray stars, converting Yohkoh/SXT solar data into corresponding focal
plane data collected by non-solar telescopes, such as ROSAT and ASCA,
from a Sun located at stellar distances.  This work presents the
application of this method to solar flares. In order to sample the
various flare conditions, we have selected a set of flares spanning
from weak to extremely intense and occurring either in simple loop
structures, or in complex regions or arcades. The method is, in general, 
the same as that used to analyse the full-disk non-flaring observations.  
However some important differences are in order: 

\begin{itemize}

\item Yohkoh SXT flare observations are performed with different 
characteristics, the so-called flare-mode, which involves mostly the 
use of two harder filters, double (full) spatial 
resolution and a much higher sampling cadence.  

\item Flares evolve on short time scales and the evolution of the 
emission measure distribution is a mostly important item of the present 
study.

\item Flares are highly localized in areas of the order of 1/1000 of
the solar hemisphere; although their luminosity is often comparable to
that of the whole corona, their evolution is mostly independent of what
happens in the rest of the corona, and we study them as self-standing
phenomena also from the stellar point of view. 

\end{itemize}

The two flare-mode filters are mostly sensitive to plasma at
temperatures above $10^7$ K, appropriate for flares; however their
sensitivity to plasma below $10^7$ K is low, and temperature
and emission measure diagnostics are not at best. Since the plasma is
typically stratified inside coronal structures, even during flares, we
may then miss contributions from relatively cool plasma components which 
may be significant, or even dominant, or coexisting with hotter plasma, 
in some pixels during the flare evolution. The presence of such a cooler 
plasma has indeed been detected in a flare observed both with flaring 
mode filters and with standard mode filters (flare 4): the emission 
measure distribution during the evolution is not modified for $T \geq 
10^7$ K but additional (lower) contributions appear for $T < 10^7$ K 
making the EM(T) less steep on the cool side of the maximum. This
should be kept in mind when discussing the shape of the EM(T)
distributions.  The slope of the cool side approaches 3/2, the typical
value expected for standard loop structures in equilibrium (see also
Peres et al. 2001), coherently to flare evolution being dominated by
plasma confined in closed magnetic structures and suggesting that,
after the initial (not bright) impulsive phases, the dynamics is
significantly less important and the flaring plasma is very close to
equilibrium conditions.

We notice that the shape and evolution of the EM(T) during flares little 
depend on the detailed geometry of the flare region, as well as on the 
flare intensity (except, of course, for the relative EM(T) height): in 
Fig.3 one can hardly distinguish flare 4, an arcade flare, and flare 8, 
very intense and complex, from the other flares with simpler geometry.  
Flare 4 and flare 8 last longer than the others, and their temperature 
decreases very slowly, but these features are not evident in Fig.3.  The 
temperature of the EM(T) maximum is also weakly dependent on the flare 
intensity: this may depend in part on the flare-mode filters used, which 
are more sensitive to plasma around that temperature, but also on the 
strong efficiency of thermal conduction ($\propto T^{5/2} \nabla T$) at 
higher and higher temperature, which implies a very large energy input 
even for a small temperature increase, to balance conductive losses to 
chromosphere.  

Significant amount of plasma at very high temperature ($\la 10^8$ K),
comparable to the peak temperatures of intense stellar flares, is
detected, in particular during flare 8. Such hot plasma, however, has
an overall low emission measure with respect to the dominant relatively
cooler plasma at $T \sim 10^7$ K. Our results show that minor
components of emission measure hotter than the dominant component
described by the single temperature fitting are practically not
detectable in low resolution spectra collected by ROSAT and ASCA during
stellar flares.  High resolution spectra collected by Chandra and
Newton may show them (Fig.6).

ROSAT and ASCA spectra synthesized from our EM(T)'s, even integrated
over long time segments and at relatively high count statistics, are
well-fitted by single isothermal components, at or around the
temperature of the EM(T) peak. Even when we include additional cooler
contributions obtained from the softer SXT filters, 1-T fit is
successful, albeit at a slightly lower best-fitting temperature.
Deviations from isothermal behavior in ROSAT and ASCA synthesized 
spectra are sometimes obtained in the decay phase because of variations 
of plasma temperature within the same time segment.  

These results can be used to interpret observations of stellar flares:
successful isothermal fittings of time-resolved spectra (such as those
mentioned in Section~\ref{sec:intro}) detect the
dominant component of a multi-component but single-peaked emission
measure distribution, and the distributions of the solar flares
presented here may be taken for reference. 
High $\chi^2$ values
during the decay may not be indicative of a multi-temperature
distribution of the flaring region, rather of the evolving temperature
of the dominant emission measure component.

We also notice that the single component fittings, both of ROSAT and
ASCA spectra, generally well evaluate the total emission measure
involved, missing only few percent of the parent emission measure.
Therefore the emission measure values obtained from the fittings are
rather reliable.

Our approach does not allow us to address exhaustively the problem of
metal abundance variations obtained to best-fit the spectra of several
stellar flares; however we made the exercise to fit the stellar-like
solar spectra letting the global metal abundance free to vary. We have
found that non-solar abundances help to improve somewhat the fitting
quality, probably because they better account for the limited broadness
of the flare EM(T)'s, but they do not seem to vary systematically (e.g.
increase first and then decrease in the decay) during each flare.
Therefore, we cannot exclude that the evidence of systematic abundance
variations requires explanations different from the one suggested by
our results.

The analysis of the ROSAT/PSPC flare simulated spectra tells us that the 
solar flares group in the X-ray flux vs hardness ratio diagram, and in 
particular in a narrow strip spanning two orders of magnitude at 
relatively high flux values and hardness ratio value between 0.2 and 
0.5. This strip is rightwards (harder), in practice completely outside,
of the region occupied by the G stars sampled by Schmitt (1997), which
however pertain to whole coronae, outside flares. This is consistent
with both flux and hardness ratio of non-flaring solar-like stellar
coronae being well below conditions of typical solar flaring regions,
which also agrees with the range of temperatures seen in non-flaring
stars vs. stellar or solar flares.

We now discuss some major implications of our results on the analysis of 
stellar coronae.  There is evidence that the emission measure 
distribution of very active stellar coronae, obtained from spectrally 
resolved XUV observations, is double-peaked; the first peak is at a few 
$10^6$ K, and the second peak above $10^7$ K (Griffiths and Jordan 
1998). This aspect is much debated and still open, but it has been 
suggested that this hot component may be due to continuous flaring 
activity (G\"udel 1997, Drake et al. 2000): the stellar surface is 
covered by active regions, flares are very frequent and their light 
curves overlap, cancelling out any variability due to the single events.  
In this framework, the present work shows that a double-peaked EM(T)
distribution is indeed obtained if one combines the EM(T) of the whole
corona to the envelope of the EM(T)'s during the flares (see Fig. 3).
The two peaks are clearly evident also when summing the EM(T)'s of
flares from 4 (M2.8) to 8 (X9) to the EM(T) of the Sun at maximum of
its cycle of activity. This seems to suggest that uninterrupted
sequences of overlapping proper flares, whichever their evolution,
would not be able to fill the gap between the two EM(T) peaks, which,
therefore, would be a permanent feature of the EM(T) of an active
star.

Indeed this second distinct maximum associated with flares is not
surprising also on the basis of theoretical modeling.  From
hydrodynamic modeling of flaring plasma confined in magnetic loops, it
has been generally found that the impulsive heating originating the
flare first causes a very fast local temperature enhancement above
$10^7$ K which propagates along the whole loop in few seconds due to
the highly efficient (even in saturated regimes) thermal conduction.
The loop plasma density (and therefore emission measure) increases more 
slowly and gradually, because determined by evaporation of plasma up 
from the chromosphere, on typical dynamic time scales (minutes). Later, 
when heating decreases, temperature and density both decay, their 
relative decay rate dictated by the rate of the heating decrease. {\it 
Models never predict that temperature does decrease and density does 
not}, which would result into an emission measure distribution 
maintaining the bell-shape, with little or no change of shape and 
shifting from high to lower temperature, thus filling the gap between 
the two EM(T) peaks.  This scenario would be at variance with both 
hydrodynamic models and observations as represented in the density-
temperature diagrams shown in Fig.1: for all flares except flare 2 the 
temperature is high and the emission measure low (upper left extremes of 
the paths) already at the beginning of the observation. In the decay the 
path is never vertical (i.e. maintaining a constant EM and decreasing 
T), but has a maximum slope (theoretically found to be less than 2, 
Jakimiec et al.  1992).  All this means that the EM(T) during flares, 
whatever its exact shape, will: 

\begin{enumerate}

\item increase starting already from $\sim 10^7$ K

\item cool but also invariably decrease, in the flare decay. The
cooling may be much slower than the decrease, if the heating is
sustained during the decay, as it seems to occur in many flares 
(Sylwester et al. 1993, Reale et al. 1997).

\end{enumerate}

These considerations and results, in our opinion, show that a major 
continuous flaring activity on a stellar corona would produce an EM(T) 
with two distinct peaks, the higher temperature one at $T \geq 10^7$ K; 
the peak will be sharper if flares are heated during their decay.  

\section{Conclusions}

This work provides a key to interpret stellar flare X-ray data in terms
of solar ones. In particular: it provides templates of stellar flare
spectra and it tells us that the single thermal components which typically
fit stellar flares low-resolution spectra are in agreement with single
peaked, relatively sharp, emission measure distributions vs
temperature.  Our simulations indicate that the best-fit temperature of
flare spectra collected with ROSAT/PSPC and ASCA/SIS corresponds to the
maximum of the flare EM(T), and that the emission measure values
obtained from the fitting well reproduce the total parent emission
measures.  This work also allows us to put solar flares in relationship
with the surface emission of stellar coronae and explains why a
continuous, moderate, flaring activity may produce a second peak in the
emission measure distribution vs temperature.

We expect more detailed information, such as the detection of minor
emission measure components at very high temperatures, from high
resolution spectra of stellar flares collected by Chandra and
XMM/Newton.

\bigskip
\bigskip
\acknowledgements{This work was supported in part by Agenzia Spaziale
Italiana and by Ministero della Universit\`a e della Ricerca
Scientifica e Tecnologica.}

\newpage

\figcaption[fig1]{Yohkoh/SXT light curves and density-temperature
diagrams for the first four of the eight flares listed in
Table~\ref{tab:fl_list}. Two light curves are shown for each flare: the
one in the Be 119 $\mu$m filter band and the other in the Al 11.4
$\mu$m filter band. The square root of the emission measure is used as
a proxy of the average density.  \label{fig:lcnt}}

\figcaption[fig2]{As in Fig.1 for the remaining four of the eight flares
listed in Table~\ref{tab:fl_list}. \label{fig:lcnt2}}

\figcaption[fig3]{Grey scale pictures of the flaring regions as imaged by
Yohokh/SXT in the Al 11.4 $\mu$m filter band. The grey scales on the
bottom left of each picture are in Data Number (DN). The position of the 
field of view on the solar disk is indicated in low right corner of each 
image.  \label{fig:frames}} 

\figcaption[fig4]{Evolution of the EM(T) during the eight selected
flares.  The emission measure distributions are sampled every $\approx 
2$ min since the beginning of the valid data.  The time sequence goes 
from lighter to darker grey shaded profiles. Examples of the EM(T)'s 
obtained in Paper II for the Sun at maximum (solid histogram) and at 
minimum (dashed histogram) of the solar cycle are also shown for 
reference (from Peres et al. 2000a). The vertical dashed line marks 
$10^7$ K.  \label{fig:emt}} 

\figcaption[fig5]{Evolution of the EM(T) during Flare 4, as in
Fig.\ref{fig:emt}, including EM contributions derived from observations
in two softer Al.1 and AlMg filter bands.  \label{fig:emt_4}}

\figcaption[fig6]{EM(T)'s yielding the maximum total emission measure of the
eight flares. Lighter to darker grey shaded profiles mark flare 1 to 8
in Table~\ref{tab:fl_list}. Reference histograms and vertical line as
in Fig.~\ref{fig:emt}.  \label{fig:maxemt}}

\figcaption[fig7]{Examples of solar X-ray spectra in the 0.2-10 keV
band synthesized from EM(T)'s at an early phase, peak and end of
Flare 2 (left panels) and at the peak and data end of Flare 8 (right
panels). The top panels show the parent spectra, the middle panels
the spectra as they would be detected by ROSAT/PSPC and the bottom 
panels as they would be detected by ASCA/SIS at stellar distances. In 
each panel of the ROSAT/PSPC and ASCA/SIS spectra, the simulated data 
(crosses) are shown together with the best--fit one--temperature MEKAL 
spectra (solid histograms) derived from the analysis (see
Table~\ref{tab:fl_asca}).  \label{fig:spectra}} 

\figcaption[fig8]{X-ray surface flux vs hardness ratio in the ROSAT/PSPC 
band obtained for various phases of the solar flares.  Each data point 
is derived from the PSPC spectrum of the EM(T)'s studied with the standard 
methods used for stellar data.  (see Table~\ref{tab:fl_asca}). 
The dashed line envelopes the stellar data reported
in Schmitt (1997).
\label{fig:rosat_flux}} 
\newpage
\begin{table}
\begin{center}
\caption[]{Selected solar flares \\}
\label{tab:fl_list}
\begin{tabular}{lcccccc}
\hline
N.&Date$^a$ & GOES Class & Duration$^b$& Start Time$^c$ & Time of EM$_{max}^d$ & End Time$^c$\\
& & & (min) & & & \\
\hline
1&920718-0505 & C5.8 & 9 & 05:08 & 05:12 & 05:18 \\
2&930210-0734 & M1.0 & 12 & 07:35 & 07:42 & 08:03 \\
3&921019-1751 & M1.1 & 20 & 17:54 & 18:00 & 18:28 \\
4&940227-0825 & M2.8 & 86 & 09:03 & 09:25 & 09:47 \\
5&911226-2133 & M4.2 & 30 & 21:35 & 21:45 & 21:55 \\
6&920206-0311 & M7.6 & 92 & 03:16 & 03:31 & 03:46\\
7&911115-2233 & X1.5 & 21 & 22:35 & 22:38 & 22:44 \\
8$^e$&921102-0231 & X9.0 & 57 & 03:04 & 03:08 & 03:35 \\
& & & & 04:42 & & 05:38 \\
& & & & 07:57 & & 08:52 \\
\hline
\end{tabular}
\end{center}
\noindent
$^a$ YYMMDD-HHMM of the flare beginning as reckoned by GOES

\noindent
$^b$ As in GOES log file

\noindent
$^c$ Start and end time of the Yohkoh/SXT observation

\noindent
$^d$ maximum of the total emission
measure obtained summing over all the pixels in each SXT frame

\noindent
$^e$ The observation is split into three segments

\end{table}

\newpage
\begin{table}
\begin{center}
\caption[]{Parameters of the selected solar flares \\}
\label{tab:fl_par}
\begin{tabular}{lcccccc}
\hline
N.&$\zeta^a$ & $\tau_{Al}^b$ & $L^c$ & Morphology$^d$ & $\log(T_{max})^e$ & $\log(EM_{max})^e$ \\
& & (sec) & $10^9$ cm & & &  \\
\hline
1& $>1.7$ & 260 & 2-3 & 1 Loop & 7.10 & 49.1 \\
2& 0.8 & 330 & 1-2 & Complex & 7.13 & 49.2 \\
3& 1.6 & 1300 & 5-10 & 1 Loop & 7.12 & 49.3 \\
4& 0.8 & 3700 & $>10$ & Arcade & 7.12 & 49.9 \\
5& 0.9 & 1100 & 5 & 2 Loops & 7.15 & 49.8 \\
6& $>1.7$ & 1800 & $>5$ & 2 Loops & 7.19 & 50.0 \\
7& 0.6 & 660 & 2-3 & 1 Loop & 7.18 & 50.4 \\
8& 0.2 & 5800 & $>5$? & Complex & 7.28 & 51.0 \\
\hline
\end{tabular}
\end{center}
\noindent

\noindent
$^a$ Slope in the n-T diagram (Reale et al. 1997)

\noindent
$^b$ Decay time of the light curve in the Al 11.4 $\mu$m filter

\noindent
$^c$ Ranges of loop half lengths estimated from the SXT images and from the
method in Reale et al. (1997).

\noindent
$^d$ From a visual analysis.

\noindent
$^e$ maximum of the weighted average temperature and total emission
measure in each frame
\end{table}

\newpage

\begin{table}
\begin{center}
\caption[]{Results of 1-T fitting of ASCA spectra of solar flares \\}
\label{tab:fl_asca}
\begin{tabular}{lccccccccc}
\hline
Flare&Dist.$^a$&Bin&Phase$^b$&Bin time range$^c$& cts&T$_{best}$&EM$_{best}$&Chan.&
$\overline{\chi}^2$\\
& (pc)&&& (sec) && ($10^7$ K) & ($10^{49}$ cm$^{-3}$) & &  \\
\hline
2&0.05&	a&	R&	0-180&		3300&	1.14&	0.54&	45&	1.4\\ 
&&	b&	R&	180-360&	6900&	1.26&	1.1&	65&	1.3\\ 
&&	c&	M&	360-540&	8200&	1.00&	1.1&	56&	1.5\\ 
&&	d&	D&	540-900&	5400&	0.77&	0.35&	43&	1.3\\ 
\hline
4&0.3&	a&	R&	0-500&		660&	1.18&	1.3&	19&	0.9\\ 
&&	b&	R$+$M&	500-1000&	2300&	1.19&	4.6&	41&	0.9\\ 
&&	c&	M$+$D&	1000-1500&	3400&	0.97&	5.9&	43&	1.4\\ 
&&	d&	D&	1500-2100&	4000&	0.95&	5.9&	43&	1.6\\ 
&&	e&	D&	2100-2700&	3900&	0.88&	5.5&	40&	1.7\\ 
\hline
6&0.2&	a&	R&	0-300&		1300&	1.15&	1.9&	29&	1.1\\ 
&&	b&	R&	300-600&	3500&	1.28&	5.5&	52&	0.9\\ 
&&	c&	M&	600-900&	4900&	1.29&	7.9&	60&	1.2\\ 
&&	d&	D&	900-1200&	4800&	1.21&	7.3&	51&	1.1\\ 
&&	e&	D&	1200-1500&	5900&	1.00&	7.7&	51&	1.8\\ 
&&	f&	D&	1500-1800&	5900&	0.89&	7.3&	47&	2.3\\ 
\hline
8&0.7&	a&	M&	0-1000&		11000&	1.50&	75&	84&	1.1\\ 
&&	b&	D&	1000-2000&	7500&	1.29&	44&	71&	1.2\\ 
&&	c&	D&	5700-7400&	4200&	1.14&	13&	49&	1.0\\ 
&&	d&	D&	7400-9100&	2900&	1.14&	8.9&	44&	1.2\\ 
&&	e&	D&	17400-20800&	400&	1.06&	0.53&	11&	0.6\\ 
\hline
4b$^d$&0.3&	a&	R&	0-500&		800&	1.1&	1.5&	23&	1.7\\ 
&&	b&	R$+$M&	500-1000&	3000&	1.0&	5.3&	42&	1.1\\ 
&&	c&	M$+$D&	1000-1500&	4100&	0.89&	6.8&	43&	2.9\\ 
&&	d&	D&	1500-2100&	5400&	0.84&	7.5&	46&	2.6\\ 
&&	e&	D&	2100-2700&	4100&	0.78&	5.5&	41&	2.8\\ 
\hline
\end{tabular}
\end{center}

\noindent
a - Distance at which the solar flare would yield the counts listed in the
6$^{\rm th}$ column

\noindent
b - R = rising phase; M = around flare maximum; D = decay phase.

\noindent
c - Reckoned since the beginning of the Yohkoh/SXT observation.

\noindent
d - Spectra including contributions from the Yohkoh/SXT softer filter bands.
\end{table}

\newpage

\begin{figure}
\figurenum{1}
\plotone{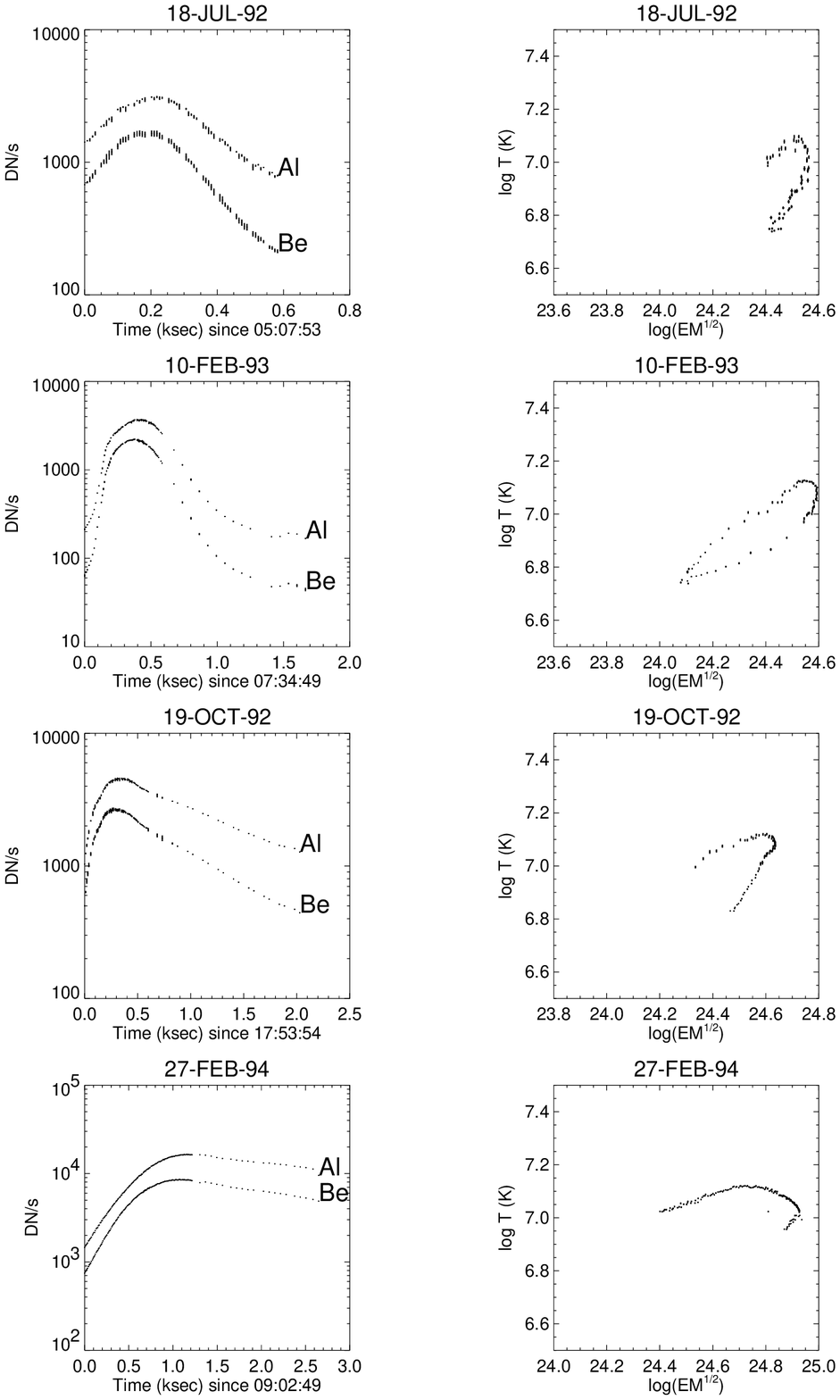}
\caption{}
\end{figure}

\begin{figure}
\figurenum{2}
\plotone{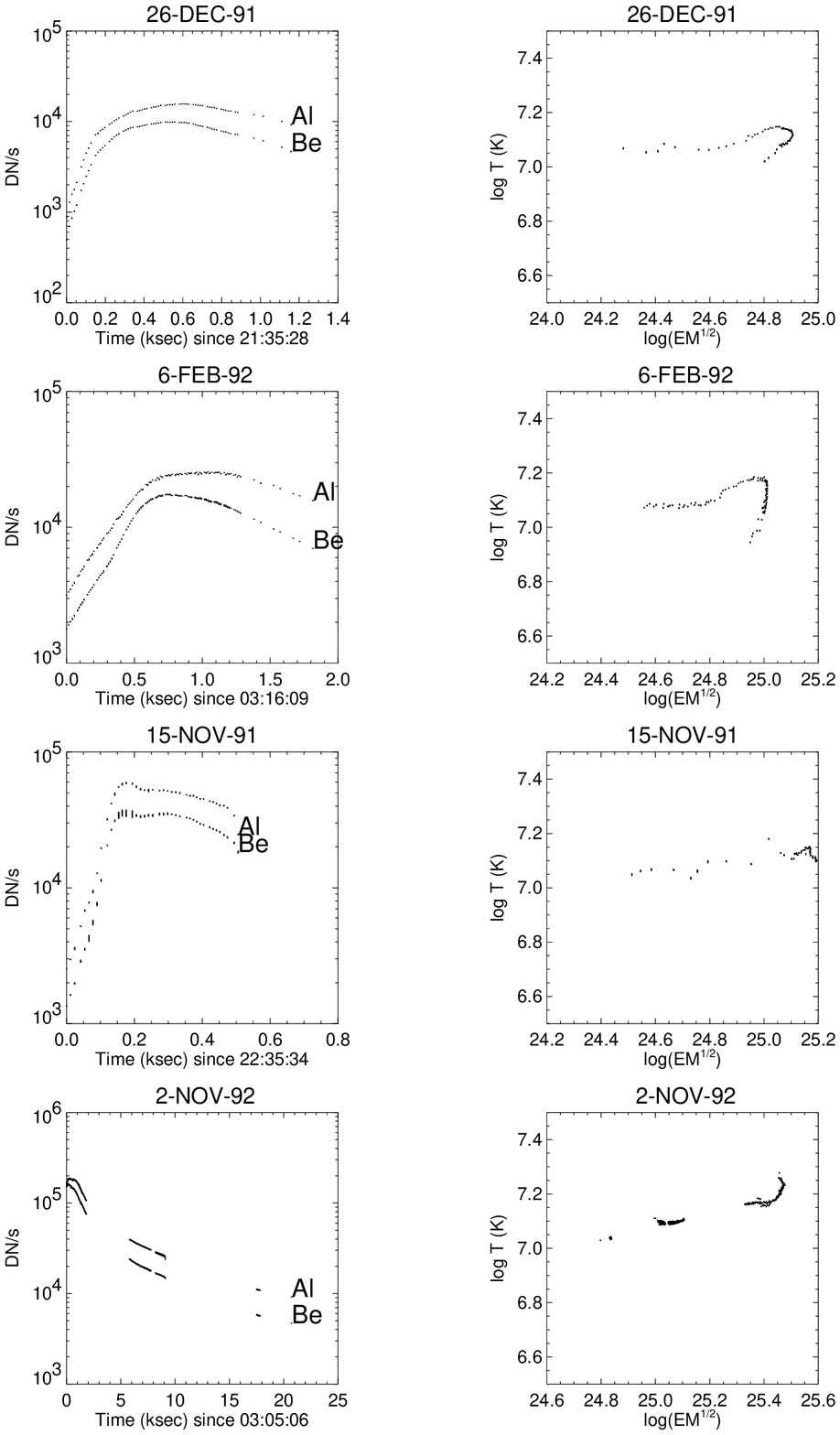}
\caption{}
\end{figure}

\begin{figure}
\figurenum{3}
\plotone{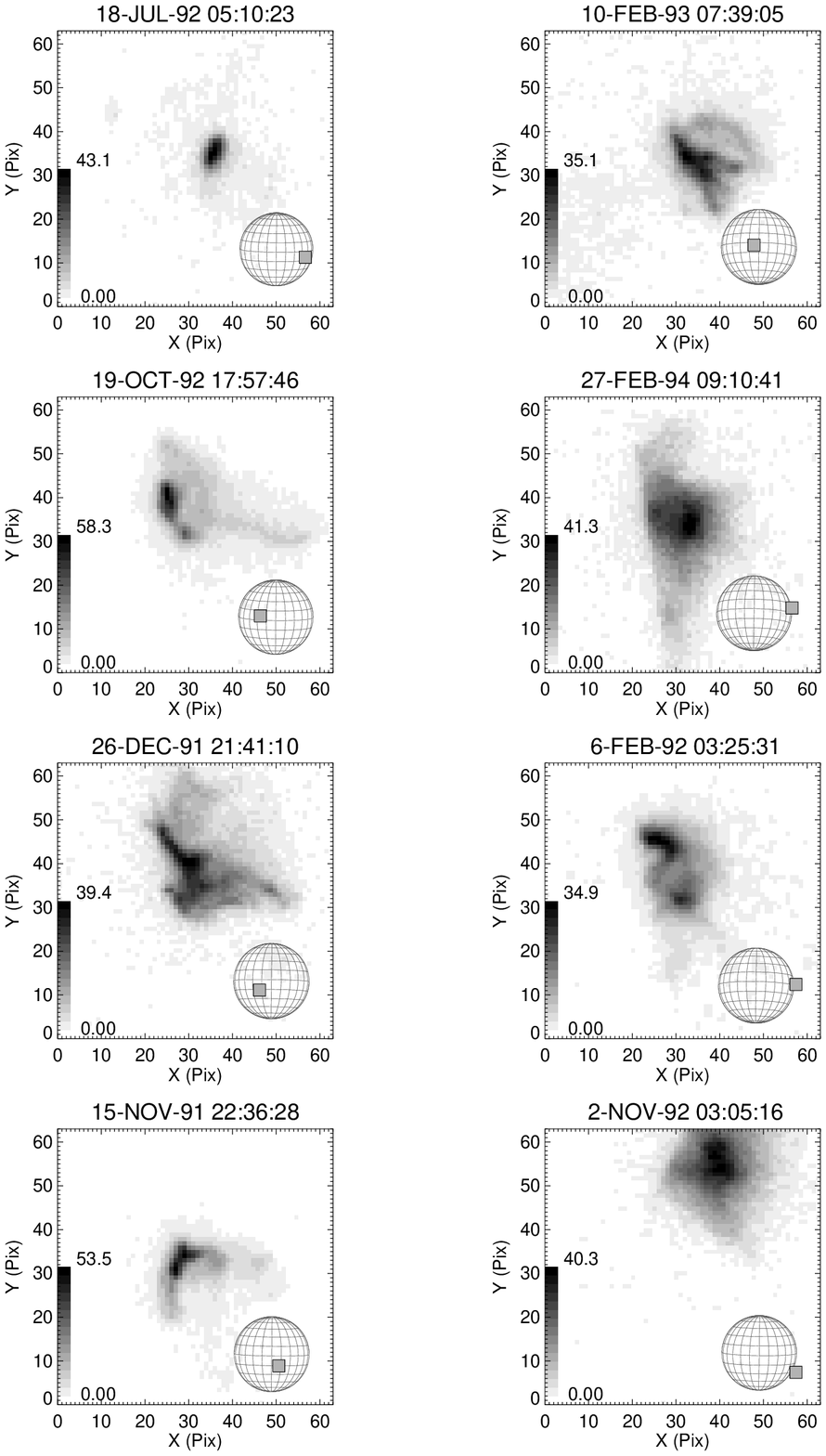}
\caption{}
\end{figure}

\begin{figure}
\figurenum{4}
\plotone{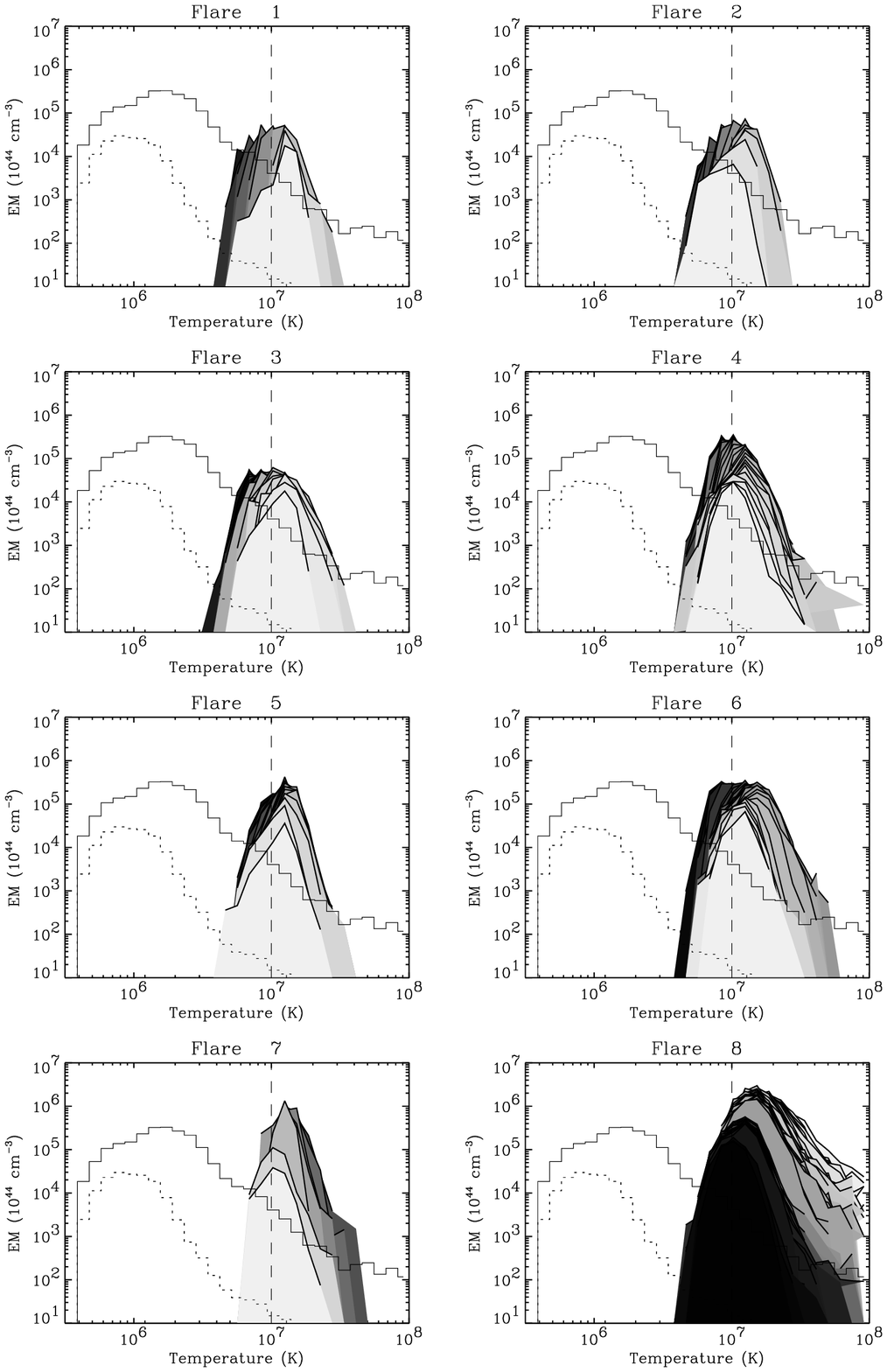}
\caption{}
\end{figure}

\begin{figure}
\figurenum{5}
\plotone{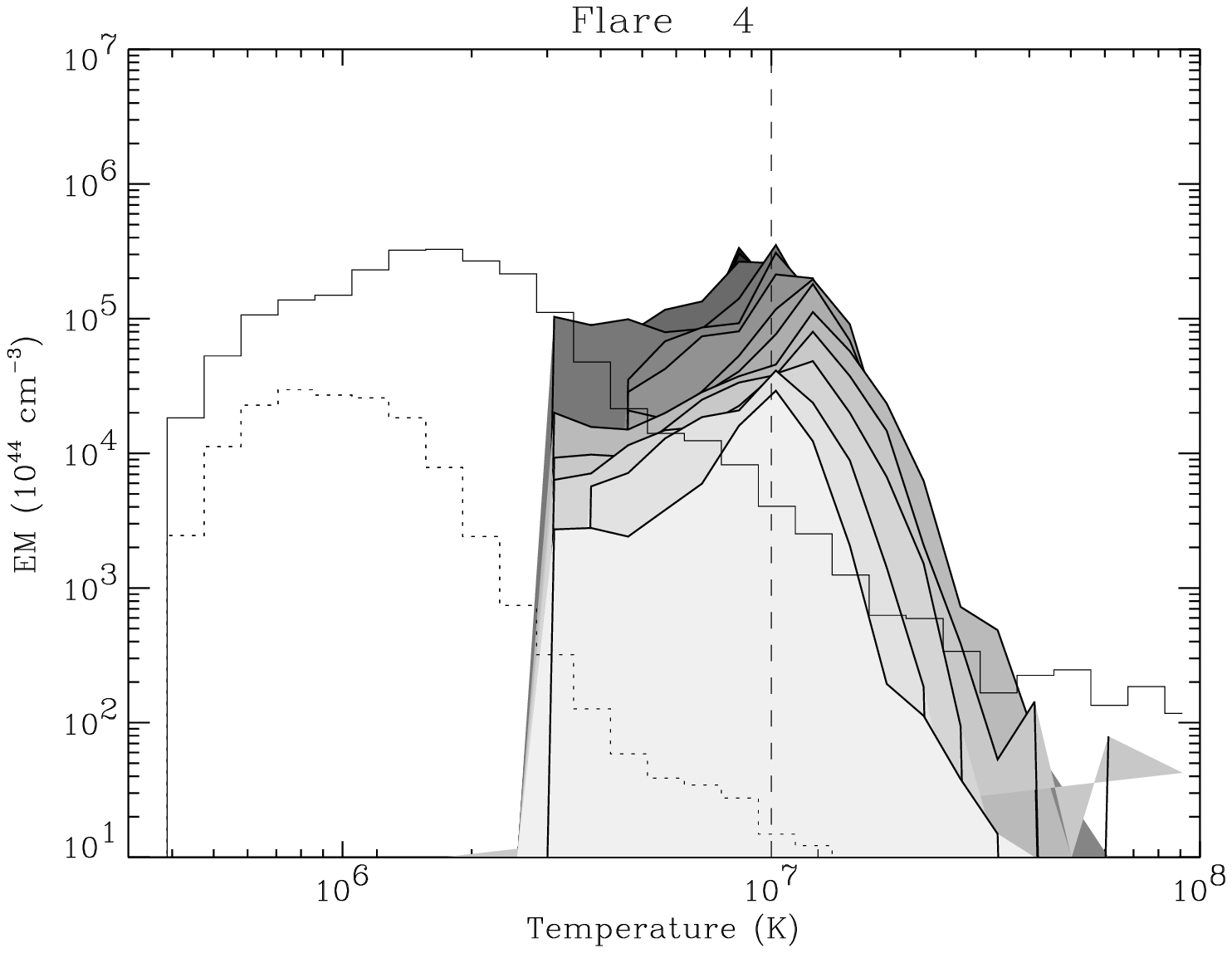}
\caption{}
\end{figure}

\begin{figure}
\figurenum{6}
\plotone{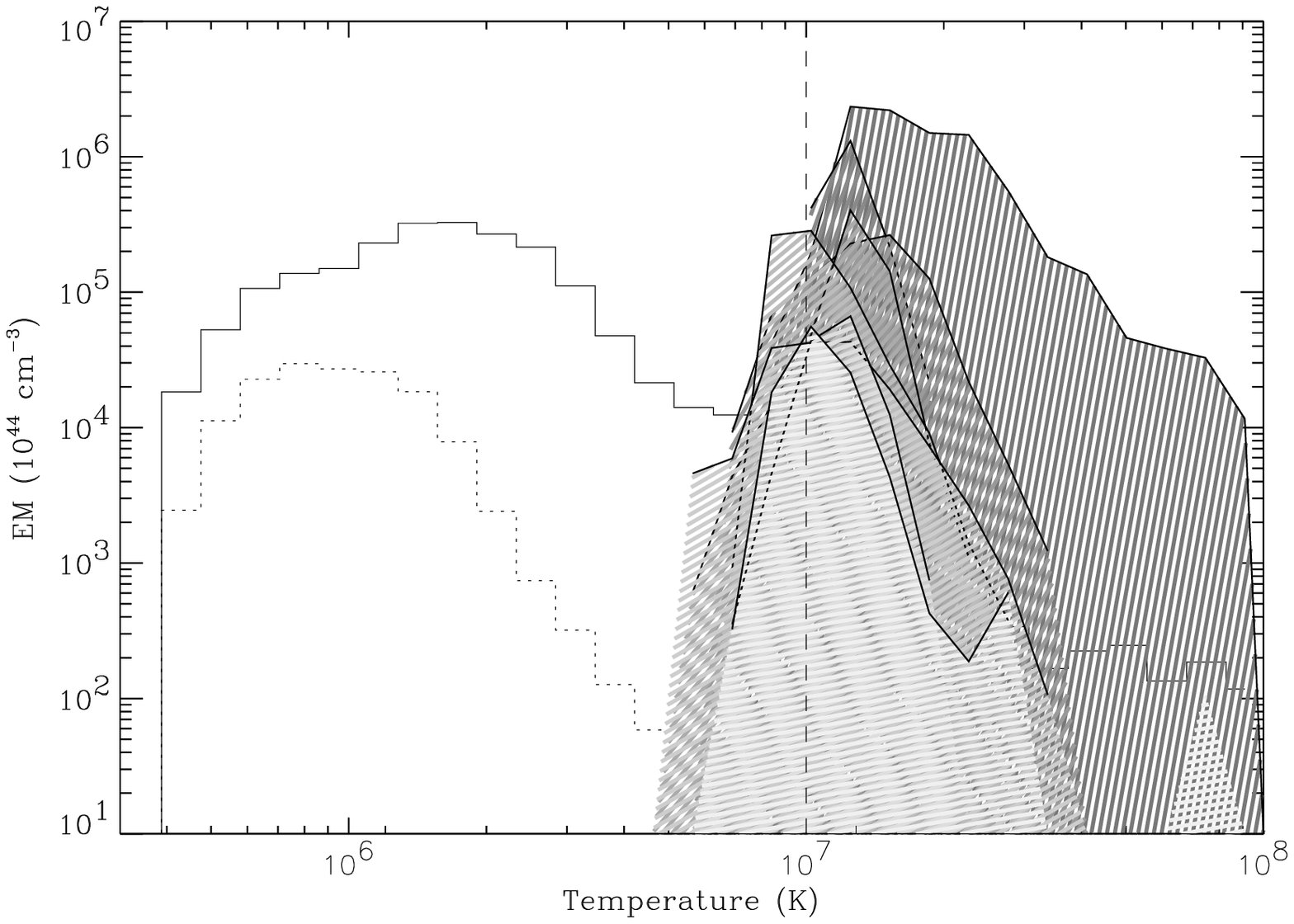}
\caption{}
\end{figure}

\begin{figure}
\figurenum{7}
\plotone{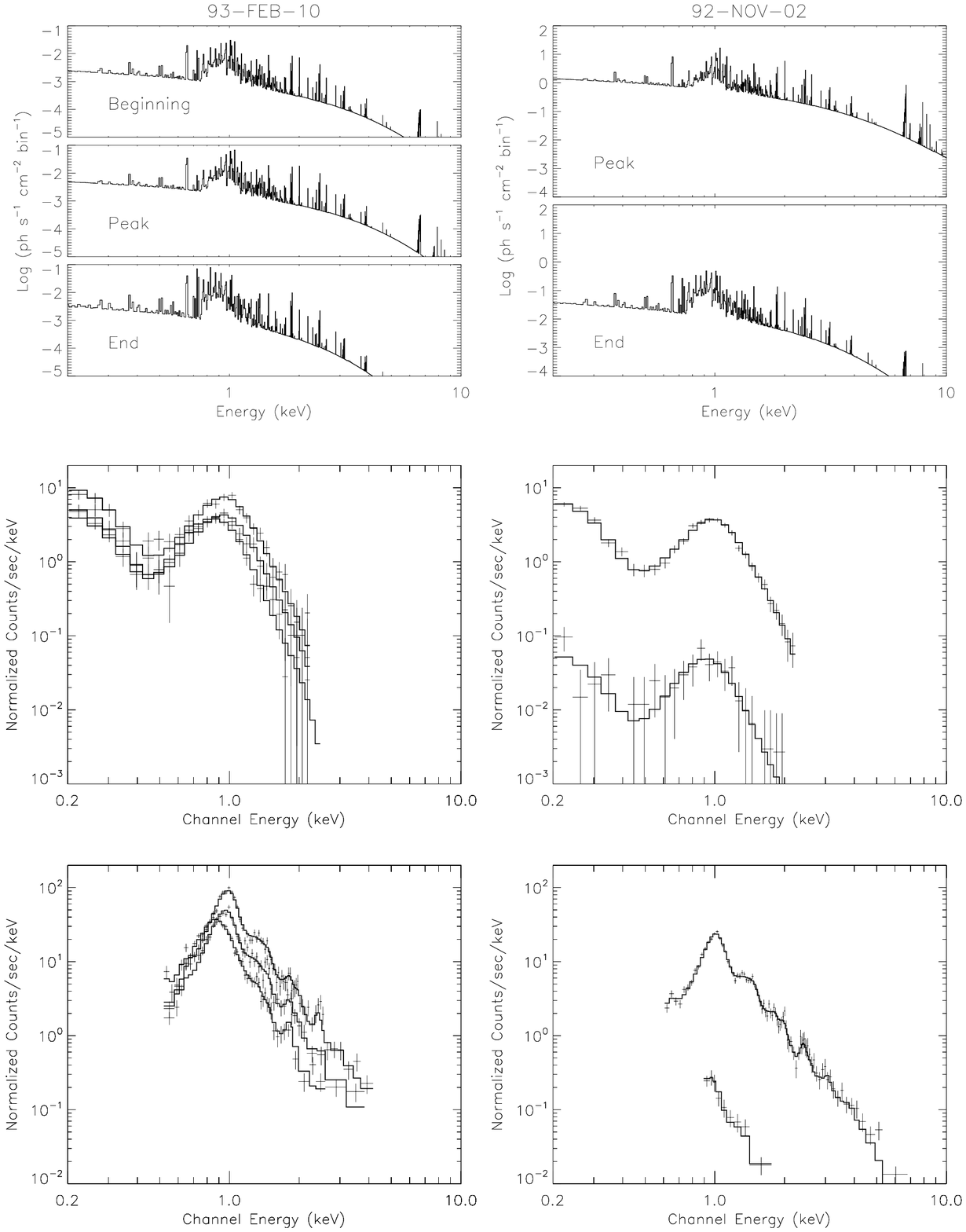}
\caption{}
\end{figure}

\begin{figure}
\figurenum{8}
\plotone{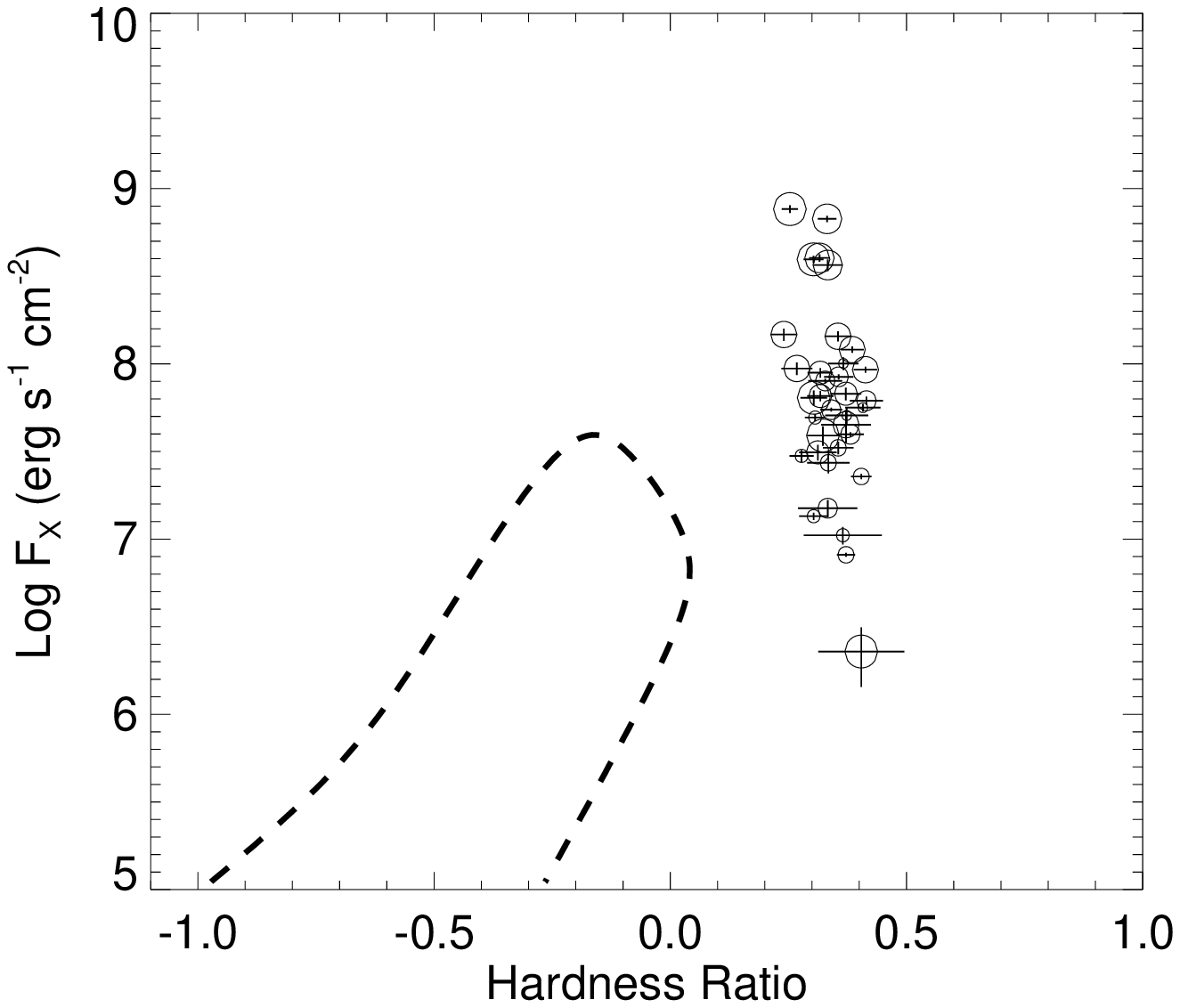}
\caption{}
\end{figure}

\end{document}